\documentstyle[prl,aps,epsfig]{revtex}         % 2 col mode
\begin{document}
\draft               % preprint mode
% 2 col mode:
\twocolumn[\hsize\textwidth\columnwidth\hsize\csname @twocolumnfalse\endcsname

\title{Spatial Resonator Solitons}

\author{V. B. Taranenko, G. Slekys, C.O. Weiss}

\address{Physikalisch-Technische Bundesanstalt \\
38116 Braunschweig, Germany}

%\date{\today{}}
\maketitle

\begin{abstract}
Spatial solitons can exist in various kinds of nonlinear optical
resonators with and without amplification. In the past years
different types of these localized structures such as vortices,
bright, dark solitons and phase solitons have been experimentally
shown to exist. Many links appear to exist to fields different
from optics, such as fluids, phase transitions or particle
physics. These spatial resonator solitons are bistable and due to
their mobility suggest schemes of information processing not
possible with the fixed bistable elements forming the basic
ingredient of traditional electronic processing. The recent
demonstration of existence and manipulation of spatial solitons
in semiconductor microresonators represents a step in the
direction of such optical parallel processing applications. We
review pattern formation and solitons in a general context, show
some proof of principle soliton experiments on slow systems, and
describe in more detail the experiments on semiconductor
resonator solitons which are aimed at applications.
\end{abstract}
 \vskip1pc ]
% BEGIN TEXT HERE
\section{Introduction: a multidisciplinary view at pattern formation and solitons}
In nonequilibrium nonlinear systems of all kinds patterns or
"structures" are known to form. Their properties seem to be
determined by a maximum dissipation principle. A system without
structure is only able to dissipate as much energy as given by
the microscopic dissipation processes. If then the system with
structure (compatible with the system equations, parameters,
boundary conditions and possibly initial conditions) has a higher
dissipation, the structure will appear spontaneously e.g. through
a modulational instability. This has been discussed at length in
relation with the Bénard convection, but is easily found to be
true in other fields.

In optics, when pumping a medium enough, then the microscopic
relaxation processes do not provide the highest possible
dissipation. As is well-known a coherent and structured and well
ordered optical laser field can build which gives the system a
much higher dissipation. If the pumping is increased, the laser
emission may convert to a pulsed instead of a continuous emission
in the form of regular or chaotic pulsing as described by the
Lorenz Model. This pulsed laser emission is again increasing the
system's capability to dissipate over the continuous emission.
Such pulsing means a modulational instability in the temporal
domain. Equally modulations or patterns appear in space, such as
a hexagonal structure of the field in the emission cross section
\cite{tag:1}.

If noise sources act, the "coherent" structures permitting the
high dissipation can appear stochastically. Examples are
appearance of vortices in wave turbulence \cite{tag:2},
vortex-antivortex generation in a noise driven Ginzburg Landau
equation \cite{tag:3}, and, evidently, phenomena such as
earthquakes and stock market crashes \cite{tag:4}. The mentioned
vortex-antivortex generation would seem to us to be a good
picture for the spontaneous appearance of particle-antiparticle
pairs out of vacuum in quantum physics and we would think that it
is also the proper model for the enigmatic phenomenon of 1/f
noise \cite{tag:5}, particularly since this description contains
the two best accepted explanations "self induced criticality"
\cite{tag:6} and the often mentioned coupled oscillator model as
special cases.

"Coherence", "structure" or "order" in space or time occurs thus
in nonequilibrium systems to maximize the energy dissipation of
the system. If one accepts such a principle (to which to our
knowledge no counter examples are known) one can see this world
as driven by a "pressure to dissipate" (where the pump evidently
is the sun). In the way in which water pressed through a porous
material or through rocks with cracks will form highly complex
flow paths through the material, one can visualize by analogy the
complex pattern formation in nature, culminating in biological
structures, as forming under the pressure to dissipate and in the
cracks and pores given by the boundary conditions in the world.

In this picture there is no need for a "plan" or "intention" to
generate the complex structures which we find developed under the
irradiation of the sun (absence of "teleological" elements).
Everything is dictated by a simple extremum principle, which is
"blind" in the sense that it can only find a nearest potential
minimum. (And the system will directly head there even if "death"
lurks in this minimum. There are good examples for such
"suicidal" systems in laser physics \cite{tag:7}.) A
"teleological" element or "planning" and an "intention" appears
only with the appearance of a brain, which can simulate the world
and find out that there are several minima of which the nearest
may not be the deepest (or that at the nearest minimum "death"
may lurk). Even then it appears that the teleological element
seems to be the exception. Even among the most complex structure,
the human species, most decisions are done unconsciously.

It is evidently one of the goals of nonlinear physics to clarify
and understand the origin of life and of biological structures.
Although the path from simple periodic patterns, appearing as a
modulational instability, and spatial solitons, to structures as
complex as biological ones would seem exceedingly long at first
glance, this may not be so at second glance.

A simple periodic pattern such as a hexagonal field structure in
the cross section of a nonlinear resonator can disintegrate into
spatial resonator solitons \cite{tag:8}. A closer examination of
these structures shows many common traits with biological
structures.

Spatial solitons in nonlinear dissipative resonators\\
1) are mobile (allowing them to move to the point in space where
they are most stable i.e. find the best "living conditions" (i.e.
they can search around for food));\\
2) they possess a "metabolism": they have internal energy flows
and spatially varying dissipation, which stabilize them; \\
3) they can "die": when the field sustaining them ("food") is
reduced, they will extinguish (die). And even a re-increase of
the field to a strength which allowed the solitons to exist
before will not reignite them, because of their bistability
(death is irreversible);\\
4) they can multiply ("self-replicate") by repeated splitting
\cite{tag:9}.\\
There are further common traits of dissipative resonator solitons
and biological structures, among them the fact that both exist
only at medium nonlinearities and not at large or small
nonlinearities (see below).

Consequently one could very well conjecture that the simplest
biological ("live") structures are not viruses (as commonly
taught) but dissipative solitons. This, by the way, shedding a
completely new light on the question of what biological life is:
life is usually ascribed to chemistry-based structures (with
carbon the central compound). Whereas if one takes the analogies
above seriously, "life" is not bound to chemistry but constitutes
itself as organizational "forms" of matter or fields. Thus the
probability of finding "live" structures is much larger than
finding "life" only in its form bound to carbon or chemistry.

These analogies of dissipative solitons with biological
structures are one of the reasons for which one might want to
study spatial solitons. Furthermore these solitons are all by
definition bistable i.e. they constitute natural carriers of
information. As opposed to the information carrying elements in
electronic information processing, they are mobile and would
therefore perhaps permit processing functions beyond the reach of
electronic computing. We have recently studied the "aggregation"
of spatial solitons and find again common traits with biology.
Considering that these "aggregations" of information carriers
constitute something like a "brain" (a collection of
information-carrying elements, bound loosely to one another,
capable of exchanging information among the elements) there is
another motivation for the study: One might hope, in the way
there are analogies between solitons and simple biological
("live") structures, to find operation principles of the brain by
studying the structure and properties of the "aggregations" of
spatial solitons.

Optics and particularly resonator optics suggest themselves for
the investigations, because of possible photonic applications
such as parallel processing, urgently awaited already e.g. in
telecom. For this reason we have conducted in the past a series
of "proof of existence"-experiments on resonator solitons as well
as of their manipulation and properties \cite{tag:10}. Presently
spatial resonator solitons in semiconductor micro-resonators are
investigated, which are most likely best suited for
applications.\\

\section{Proof of existence experiments of resonator solitons with slow materials}

Spatial resonator solitons are self-formed localized structures,
which are free to move around. We have investigated several types:

\begin{figure}[htbf] \epsfxsize=60mm
\centerline{\epsfbox{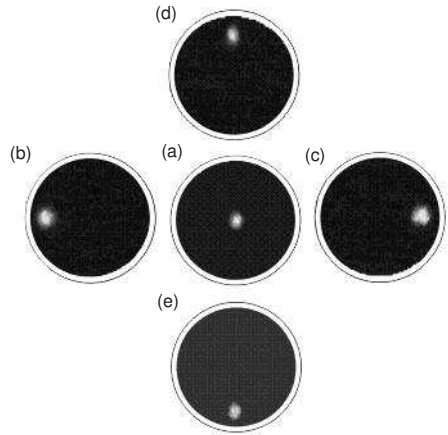}} \vspace{0.7cm} \caption{In a
laser with internal nonlinear absorber (bacterio-rhodopsin) a
soliton can be written in arbitrary places in the resonator cross
section. The resonator used is degenerate for all transverse modes
to allow any arbitrary field configuration to be resonant.}
\end{figure}

Vortices occur in resonators with a gain medium without phase
preference, such as laser resonators. They are helical defects in
the wavefront and possess therefore, strictly speaking,
"tristability": right- as well as left-handed vortices coexist
with the wave without defect. We have given an overview of laser
vortices in \cite{tag:10} including examples of their dynamics in
Internet films \cite{tag:11}. Vortices were until now not studied
with regard to their usefulness as binary information carriers
(except for some basic experiments relating to pattern
recognition with lasers \cite{tag:12}).

\begin{figure}[htbf] \epsfxsize=40mm
\centerline{\epsfbox{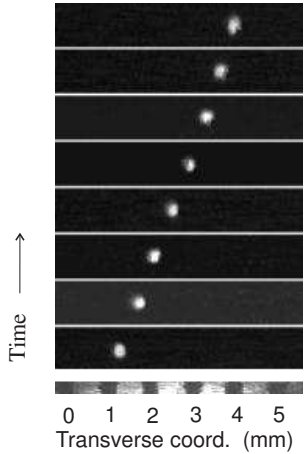}} \vspace{0.7cm} \caption{A soliton
can be made to move across the resonator cross section by a phase
gradient in the resonator.}
\end{figure}

\begin{figure}[htbf]
\epsfxsize=40mm \centerline{\epsfbox{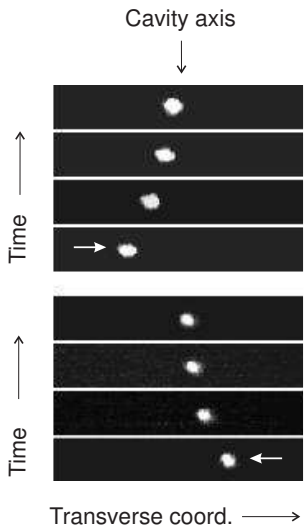}} \vspace{0.7cm}
\caption{In a resonator with a "phase trough" at the center a
soliton is drawn from all sides to the resonator axis and is
trapped there.}
\end{figure}

Bright solitons are more reminiscent of the original idea of
optical information processing namely the use of optically
bistable resonators. They have therefore been considered more
closely for such tasks. In order to convince oneself that such
bright solitons in bistable resonators exist generally, we
conducted experiments with nonlinear media with slow response -
for the purpose of being able to observe the phenomena (which
entail 2D space-time dynamics) on a convenient time scale. As
opposed to semiconductors which have characteristic times of ps
to ns, e.g. photorefractive media and nonlinear absorbers like
bacterio-rhodopsine \cite{tag:13} have time constants of 10 ms to
1 s, quite suitable for recording with ordinary video equipment.
In general a resonator containing only such a non-linear absorber
is too lossy for bistability. Therefore we used resonators
containing additionally gain elements to compensate for the
losses. A system like a laser with nonlinear absorber results
which is well suited to study these transverse effects such as
spatial modulational instabilities, patterns and solitons.

\begin{figure}[htbf]
\epsfxsize=75mm \centerline{\epsfbox{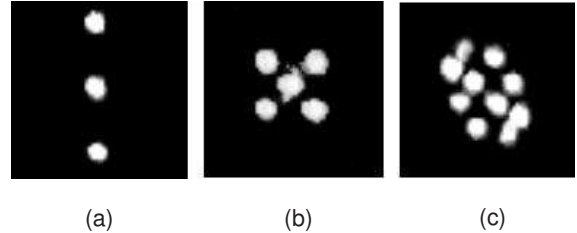}} \vspace{0.4cm}
\caption{Larger number of stationary bright solitons coexisting.}
\end{figure}

The first of such experiments yielded evidence of the existence of
spatial solitons \cite{tag:14}. In refined experiments we showed
the existence, the bistability and the manipulations of bright
soliton in this system \cite{tag:15}. FIG.~1 shows that the
solitons can be "ignited" anywhere in the cross section of the
resonator. FIG.~2 shows the motion of a bright soliton in a phase
gradient, and FIG.~3 shows the "trapping" of a bright soliton in a
"phase-trough". These are functions as they are required for
applications. FIG.~4 shows that large numbers of such bright
solitons can exist at the same time \cite{tag:9}, equally as
necessary for technical applications.

\begin{figure}[htbf]
\epsfxsize=45mm \centerline{\epsfbox{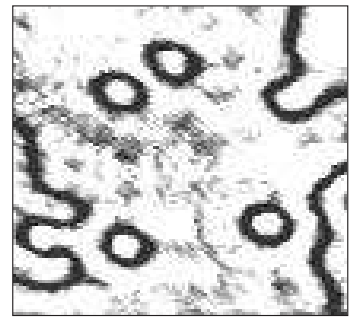}} \vspace{0.7cm}
\caption{Phase solitons coexisting with non stationary phase
fronts. At the fronts and surrounding the soliton the phase of the
field changes by 180 degrees producing by interference the dark
ring surrounding the solitons and the dark fronts at the
phase-domain boundaries.}
\end{figure}

As an interesting feature of these solitons it was found
\cite{tag:16} that these solitons have quantized velocity
magnitudes (including velocity magnitude zero). Although the
velocity magnitudes of such solitons are fixed, their direction
of motion is completely free, and varies e.g. under the action of
noise.

It may be mentioned that in resonators with phase-sensitive gain,
such as gain supplied by 4-wave-mixing or degenerate parametric
mixing, solitons which concern only the phase structure of the
field (phase-solitons) were predicted \cite{tag:17} and
experimentally observed \cite{tag:18}. See FIG.~5.\\

\section{Semiconductor Resonator Solitons }

To make bright solitons suitable for technical applications, fast
materials have to be used. We found that the optimum system
concerning speed of response and nonlinearity is the
semiconductor micro-resonator. This is a resonator of $\sim$ 1
wavelength length which provides the shortest conceivable
resonator response time of 100 fs - 1 ps. The finesse of these
resonators has to be around 100 - 200 to provide bistability
together with a nonlinear medium. For the latter a semiconductor
slice of the length of the resonator ($\sim$ 1$\lambda$) is
suitable, and we stress that the response time of semiconductor
material is well matched to that of the resonator. Importantly:
it makes no sense to use "faster" materials than semiconductors
in such resonators. Since the small length of the resonator
provides a response time which is at the conceivable limit for
optical resonators, a faster nonlinear material does not decrease
the system response time. However, since higher speed comes at the
cost of a smaller nonlinearity, a faster material would require
unnecessarily high light intensities. Thus the $\sim$ 1$\lambda$
microresonator combined with a semiconductor material as the
nonlinear ingredient represents the optimum in terms of response
time and required light intensity. This structure is incidentally
the structure of VCSELs (vertical cavity surface emitting laser)
which during the last years have already been developed to a
certain degree of perfection. For the existence of spatial
solitons in such resonators absorbing ("passive") as well as
population-inverted ("active") material is suitable. Thus a
direct connection with VCSEL technology exists.

FIG.~6 shows the structure of the nonlinear resonator. Its length
is $\sim$ 1$\lambda$ while the transverse size is typically 5 cm.
The short length and wide area of this microresonator permit only
one longitudinal mode (FIG.~7~(a)) while allowing an enormous
number of transverse modes so that a very large number of spatial
solitons can coexist. The resonator is obviously of the plane
mirror type, implying frequency degeneracy of all transverse
modes and thus allowing arbitrary field patterns to be resonant
inside the resonator. This is another prerequirement for
existence and manipulability of spatial solitons.

\begin{figure}[htbf]
\epsfxsize=80mm \centerline{\epsfbox{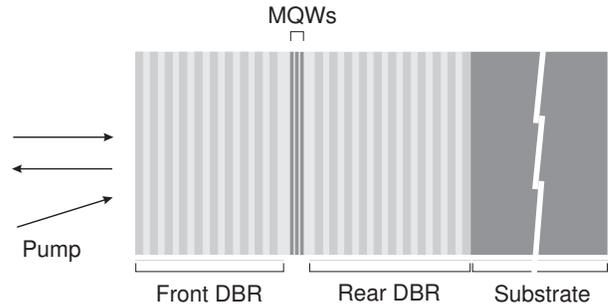}} \vspace{0.7cm}
\caption{Schematic of semiconductor microresonator consisting of
two plane distributed Bragg reflectors (DBR) and multiple quantum
wells (MQW). Observation is in reflection.}
\end{figure}

\begin{figure}[htbf]
\epsfxsize=85mm \centerline{\epsfbox{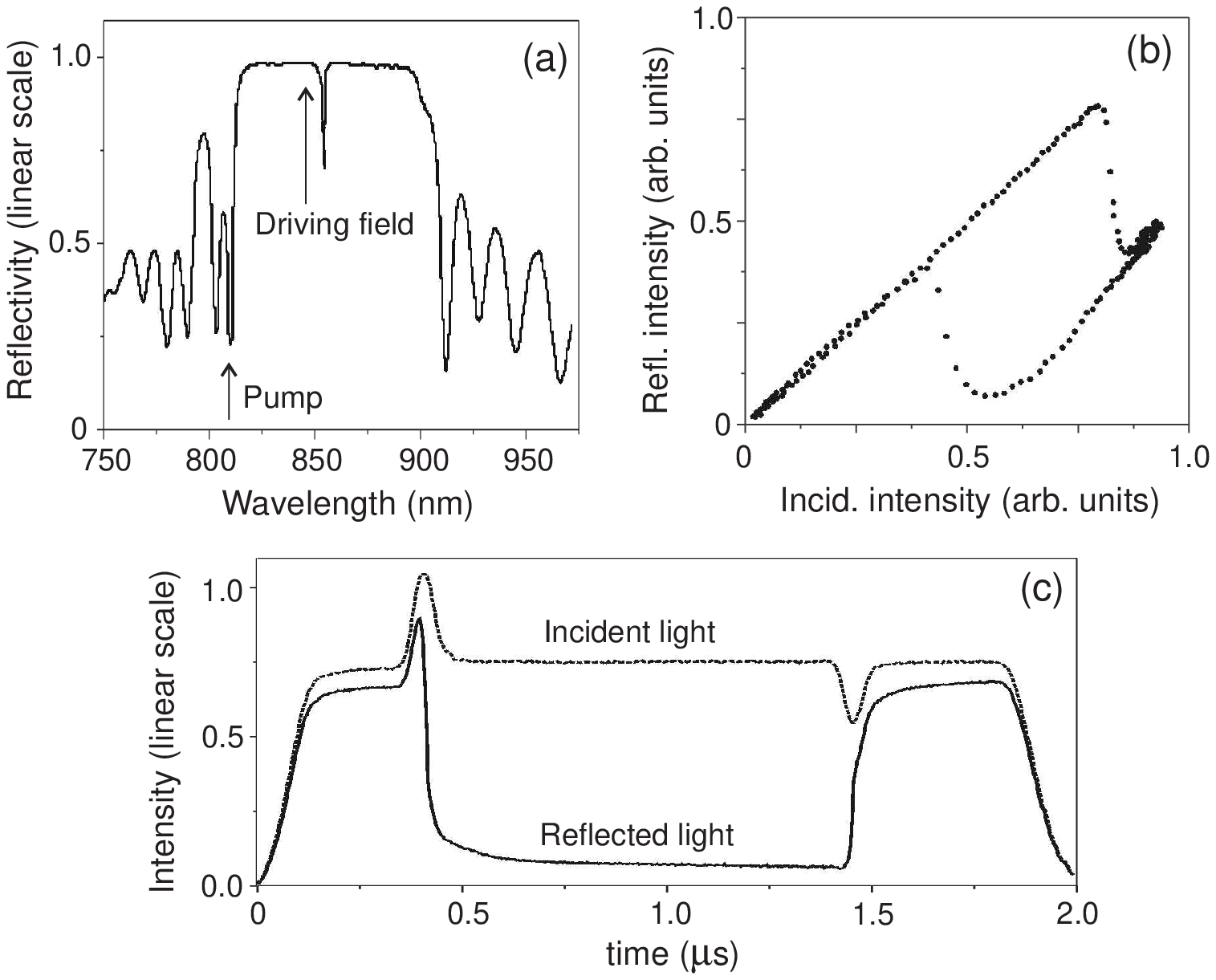}} \vspace{0.7cm}
\caption{Semiconductor microresonator reflectance spectrum (a),
typical bistability loop in reflection (b) and dynamics of
switching (c). Arrows mark the driving field that is detuned from
the resonator resonance and the pump field that is tuned to be
coupled into the resonator through one of the short-wavelength
interference notches of the resonator reflectance spectrum. (c)
shows that by short positive (negative) pulses the resonator can
be switched from one branch of bistability loop to the other.}
\end{figure}

The resonator soliton existence is closely linked with the plane
wave resonator bistability (FIG.~7~(b)) caused by longitudinal
nonlinear effects: the nonlinear changes of the resonator length
(due to nonlinear refraction changes) and finesse (due to
nonlinear absorption changes) \cite{tag:19}. The longitudinal
nonlinear effects combined with transverse nonlinear effects
(such as self-focusing) can balance diffraction and form resonator
solitons. Generally these nonlinear effects can cooperate or act
oppositely, with the consequence of reduced soliton stability in
the latter case.

\subsection{Model and numerical analysis}

As a guide for the experiments we use a phenomenological model of
a driven wide area multiquantum well (MQW)-semiconductor
microresonator similar to \cite{tag:20,tag:21}. The optical field
$A$ inside the resonator is described in the mean-field
approximation \cite{tag:22}. The driving incident field {$A_{\rm
in}$} is assumed to be a stationary plane wave. Nonlinear
absorption and refractive index changes induced by the
intracavity field in the vicinity of the MQW-structure band edge
are assumed to be proportional to the carrier density $N$
(normalized to the saturation carrier density). The equation of
motion for $N$ includes optical pumping $P$, carrier recombination
and diffusion. The resulting coupled equations describing the
spatio-temporal dynamics of $A$ and $N$ have the form:

\begin{eqnarray}
\cr {\partial A}/{\partial t}=A_{\rm in}-\sqrt{T}A\{[1+C{\rm
Im}(\alpha)(1-N)]+\cr{+i(\theta-C{\rm
Re}(\alpha)N-\nabla^{2}_{\bot})\}}\,, \cr\cr{\partial N}/{\partial
t}=P-\gamma[N-|A|^2(1-N)-d\nabla^{2}_{\bot}N]\,\,,
\end{eqnarray}

where $C$ is the saturable absorption scaled to the resonator
transmission $T$ ($T$ is assumed to be small since the mirror
reflectivity is typically $\geq$ 0.995). ${\rm Im}(\alpha)(1-N)$
and ${\rm Re}(\alpha)N$ describe the absorptive and refractive
nonlinearities, respectively. $\theta$ is the detuning of the
driving field from the resonator resonance. $\gamma$ is the photon
lifetime in the resonator normalized to the carrier recombination
time. $d$ is the diffusion coefficient scaled to the diffraction
coefficient, and $\nabla^{2}_{\bot}$ is the transverse Laplacian.

\begin{figure}[htbf]
\epsfxsize=70mm \centerline{\epsfbox{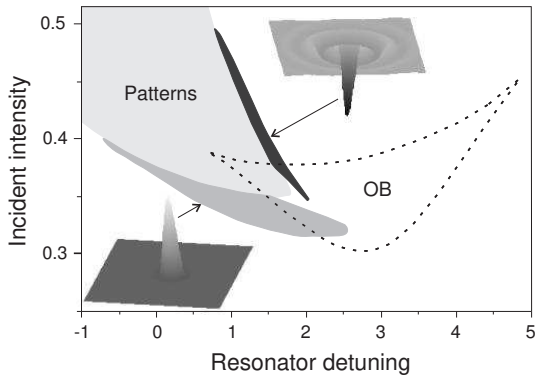}} \vspace{0.7cm}
\caption{Numerical solutions of Eq.(1) for unpumped and mixed
absorptive/self-focusing case. Area limited by dashed lines is
optical bistability domain for plane waves. Shaded areas are
domains of stability for bright/dark solitons and patterns. Insets
are bright and dark solitons in 3D representation.}
\end{figure}

Linear effects in the resonator are spreading of light by
diffraction, and carrier diffusion (terms with
$\nabla^{2}_{\bot}$ in (1)). The material nonlinearity that can
balance this linear spreading can do this in various ways. It has
a real (refractive) and imaginary (dissipative) part and can act
longitudinally and transversely. The nonlinear changes of the
resonator finesse (due to nonlinear absorption change) and length
(due to nonlinear refractive index change) constitute
longitudinal nonlinear effects, also known under the name
\emph{nonlinear resonance} \cite{tag:17}. The transverse effects
of the nonlinear refractive index can be self-focusing (favorable
for bright and unfavorable for dark solitons) and self-defocusing
(favorable for dark and unfavorable for bright solitons).
Absorption (or gain) saturation (bleaching) leading to
\emph{nonlinear gain guiding} (in laser parlance), is a transverse
effect. Longitudinal and transverse effects can work oppositely,
or cooperate.

\begin{figure}[htbf]
\epsfxsize=70mm \centerline{\epsfbox{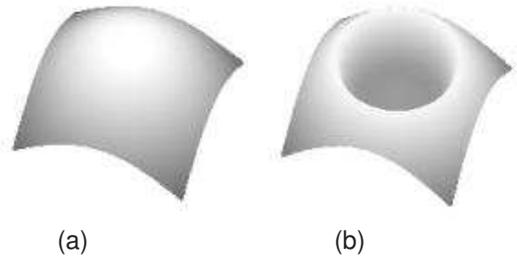}} \vspace{0.7cm}
\caption{Maxwellian switched domain in reflection (b) numerically
calculated for parameters corresponding to the plane-wave OB area
in FIG.~8 and for Gaussian profile (a) of the driving beam.}
\end{figure}

\begin{figure}[htbf]
\epsfxsize=70mm \centerline{\epsfbox{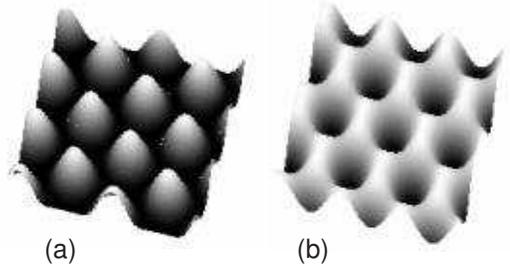}} \vspace{0.7cm}
\caption{Bright- (a) and dark- (b) spot-hexagons numerically
calculated for parameters corresponding to the Pattern existence
domain of FIG.~8. Driving light intensity increases from (a) to
(b).}
\end{figure}

There are two principal external control parameters: the driving
field intensity $|A_{\rm in}|^2$ and the resonator detuning
$\theta$. FIG. 8 shows typical existence domains (in coordinates
$\theta$, $|A_{\rm in}|^2$) for all possible structures
(patterns, bright and dark spatial solitons) and plane-wave
bistability as calculated from (1) for the case of a mixed
absorptive/dispersive nonlinearity. At large resonator detuning
the intracavity field is transversely homogeneous and stable. If
in this case the driving beam has a Gaussian profile (FIG.~9~(a))
and its amplitude in the maximum exceeds the switching-on
threshold then the part of the beam cross section which is
limited by so called Maxwellian intensity \cite{tag:23} is
switched, thus forming a dark switched domain in reflection
(FIG.~9~(b)).

For small resonator detuning and for driving intensities not
quite sufficient for reaching the resonance condition for the
whole resonator area, the system "chooses" to distribute the
light intensity in the resonator in isolated spots where the
intensity is then high/low enough to reach the resonance
condition, thus forming bright/dark patterns (FIG.~10). Instead
of saying "the system chooses" one would more mathematically
express this by describing it as a modulational instability. The
detuned plane wave field without spatial structure with intensity
insufficient to reach the resonance condition is unstable against
structured solutions. According to our numerical solutions of (1)
a large number of such structured solutions coexist and are
stable (see e.g. patterns in FIG.~18).

The bright/dark soliton structures (inset in FIG.~8) can be
interpreted as small circular switching fronts, connecting two
stable states: the low transmission and the high transmission
state. Such a front can in 2D surround a domain of one state.
When this domain is comparable in diameter to the "thickness" of
the front, then each piece of the front interacts with the piece
on the opposite side of the circular small domain, which can
lead, particularly if the system is not far from a modulational
instability (see FIG.~8, Patterns), to a stabilisation of the
diameter of the small domain. In this case the small domain is an
isolated self-trapped structure or a dissipative resonator
soliton.

\subsection{Experimental arrangement}

FIG.~11 shows the optical arrangement for the semiconductor
soliton experiments, and in particular for their switching on or
off. The semiconductor microresonator consists of a MQW
(GaAs/AlGaAs or GaInAs/GaPAs) structure sandwiched between
high-reflectivity ($\geq$ 0.995) DBR-mirrors (FIG.~6) used at room
temperature. The microresonator structures were grown on GaAs
substrates by a molecular beam epitaxy technique that allows
growing MQW structures with small radial layer thickness
variation. The best sample used in our experiments has only
$\sim$ 0.3 nm/mm variation of the resonator resonance wavelength
over the sample cross section of 5 cm diameter.

The driving light beam was generated by either a tunable (in the
range 750-950 nm) Ti:Sa laser or a single-mode laser diode
($\sim$ 854 nm), both emitting continuously. For experimental
convenience and to limit thermal effects, the experiments are
performed within a few microseconds, by admitting the light
through an acusto-optical modulator. A laser beam of suitable
wavelength is focused onto the microresonator surface in a spot
of $\sim$ 50 $\mu$m diameter, thus providing quite large Fresnel
number ($\geq$ 100).

\begin{figure}[htbf]
\epsfxsize=85mm \centerline{\epsfbox{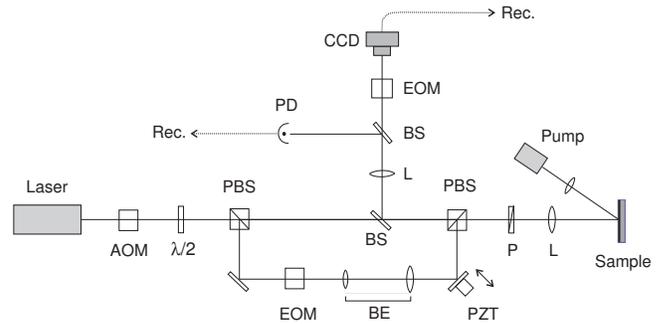}} \vspace{0.7cm}
\caption{Experimental setup. Laser: Ti:Sa (or diode) laser, AOM:
acousto-optic modulator, $\lambda$/2: halfwave plate, PBS:
polarization beam spliters, EOM: electro-optical amplitude
modulators, BE: beam expander, PZT: piezo-electric transducer, P:
polarizer, L: lenses, BS: beam splitters, PD: photodiode. In some
cases an optical pump is used ("Pump"), see text.}
\end{figure}

Part of the laser light is split away from the driving beam and
is superimposed with the main beam in a Mach-Zehnder
interferometer arrangement, to serve as a writing/erasing
(address) beam. This beam is tightly focused and directed to some
particular location in the illuminated area to create or destroy
a spatial soliton. The switching light is opened for a few
nanoseconds using an electro-optic modulator. For the case of
incoherent switching the polarization of the address beam is
perpendicular to that of the main beam to avoid interference. For
the case of coherent switching the polarizations are parallel and
a phase control of the switching field is always needed: for
switching on as well as switching off a soliton. One of the
interferometer mirrors can be moved by a piezo-electric element
to control the phase difference between the background light and
the address light.

Optical pumping of the MQW-structures was done by a multi-mode
laser diode or a single-mode Ti:Sa laser. To couple the pump
light into the microresonator the pump laser wavelength was tuned
into the short wavelength reflection minimum as shown in
FIG.~7~(a).

The observations are done in reflection (because the GaAs
substrate is opaque) by a CCD camera combined with a fast shutter
(another electro-optic modulator), which permits to take
nanosecond snapshots at a given time, of the illuminated area on
the resonator sample. Recording movies on the nanosecond time
scale is also possible. To follow the intensity in time in
certain points (e.g. at the location of a soliton) a fast
photodiode can be imaged onto arbitrary locations within the
illuminated area.

\subsection{Results and discussions}

Switched (dark) domains exist for small negative resonator
detuning in the vicinity of the absorption band edge. The
switching front surrounding these domains is defined by the
spatial profile of the driving beam. The switching front is
located where the driving intensity equals the Maxwellian
intensity. FIG.~12 shows snapshots of various switched domains
demonstrating that they have the same shapes as the driving
beams. When changing the intensity of the driving beam in time
the fronts around the switched domains follow obviously an
equiintensity contour of the incident light (FIG.~13).

\begin{figure}[htbf]
\epsfxsize=70mm \centerline{\epsfbox{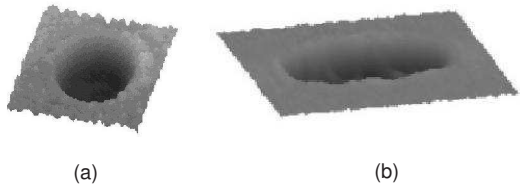}} \vspace{0.7cm}
\caption{Dark switched domains observed at small (negative)
resonator detuning for round (a) and oval (b) driving beam
shapes.}
\end{figure}

\begin{figure}[htbf]
\epsfxsize=60mm \centerline{\epsfbox{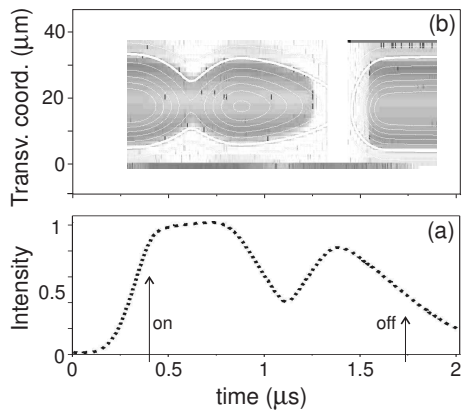}} \vspace{0.7cm}
\caption{Incident driving light intensity, center of Gaussian beam
(a) and reflectivity of sample (on diameter of circular driving
beam cross section) along with equiintensity contours of incident
light (b) demonstrating that borders of dark switched area follow
one of the equiintensity contours (heavy line).}
\end{figure}

In contrast to these dark Maxwellian switched domains, spatial
patterns and solitons are self-sustained objects independent of
boundary conditions or a beam profile and can be both bright and
dark. FIG.~14 shows snapshots of bright and dark small ($\sim$ 10
$\mu$m) round spots (solitons), at large (negative) resonator
detuning, whose shape/size is independent of the shape/intensity
of the driving beam \cite{tag:24}. This independence of boundary
conditions allows to distinguish between patterns and solitons on
the one side and Maxwellian switched domains on the other side.

\begin{figure}[htbf]
\epsfxsize=80mm \centerline{\epsfbox{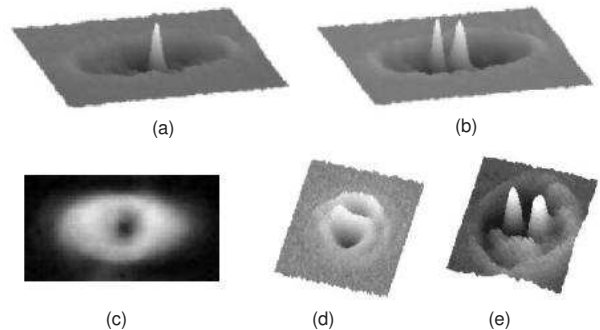}} \vspace{0.7cm}
\caption{Bright- and dark-spot switched structures observed
experimentally at large (negative) resonator detuning. The
independence of the structures of the driving beam shape shows
that the structures are self-localized (patterns or spatial
solitons).}
\end{figure}

\begin{figure}[htbf]
\epsfxsize=60mm \centerline{\epsfbox{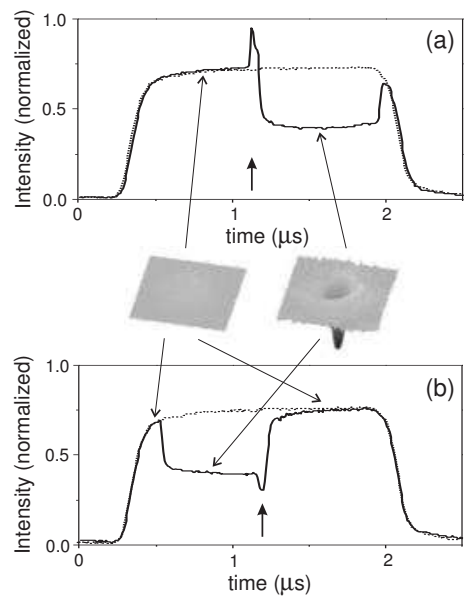}} \vspace{0.7cm}
\caption{Recording of coherent switching-on (a) and switching-off
(b) of a bright soliton. Heavy arrows mark the application of
switching pulses. Dotted traces: incident intensity; solid traces:
reflected intensity at center of soliton. The insets show
intensity snapshots, namely unswitched state (left) and soliton
(right). Details see text.}
\end{figure}

Resonator solitons can be written and erased by focused optical
(coherent) pulses independent of other bright (dark) spots
(solitons).

FIG.~15 shows how solitons can be switched by light coherent with
the background light \cite{tag:25}. FIG.~15~(a) shows switching a
bright soliton on. The driving light intensity is chosen slightly
below the spontaneous switching threshold. At ${t}$ $\approx$ 1.2
$\mu$s the writing pulse is applied. It is in phase with the
driving light, as visible from the constructive interference. A
bright soliton results, showing up in the reflected intensity
time trace as a strong reduction of the intensity. FIG.~15~(b)
shows switching a soliton off. The driving light is increased to
a level where a soliton is formed spontaneously. The address
pulse is then applied in counterphase to the driving light, as
visible from the destructive interference. The soliton then
disappears, showing up in the reflected intensity time trace as
reversion to the incident intensity value. The FIG.~15 insets
show 2D snapshots before and after the switching pulses for
clarity.

Thus depending on conditions we find switched domains, patterns
and bright and dark resonator solitons. Periodic patterns can be
distinguished from collections of solitons by the mutual
independence of the latter.

To find the most stable resonator solitons for applications one
can play with the nonlinear (absorptive/dispersive) material
response by choice of the driving field wavelength and intensity,
with the resonator detuning, and finally with the carrier
population inversion (when using pumping). We recall that all
nonlinearities change their sign at transparency i.e. at the
point where the valence- and the conduction band populations are
equal. Going from below transparency (absorption) to above
transparency (population inversion, producing light
amplification), nonlinear absorption changes to nonlinear gain,
self-focusing changes to self-defocusing and vice versa, and
decrease of optical resonator length with intensity changes to
increase (and vice versa). The population of the bands can be
controlled by pumping ($P$ in (1)) i.e. transferring electrons
from the valence band to the conduction band. We do this by
optical excitation \cite{tag:26}, with radiation of a wavelength
shorter than the band edge wavelength. If the structures were
suited to support electrical currents (i.e. if it were a real
VCSEL-structure) pumping could evidently be effected by
electrical excitation.

\subsubsection{Illumination below bandgap (defocusing nonlinearity)}

Working well below the bandgap (in the dispersive/defocusing
limit) with the driving field wavelength $\sim$ 30 nm longer than
the band edge wavelength we observe spontaneous formation of
hexagonal patterns (FIG.~16). The hexagon period scales linearly
with  $\theta^{-1/2}$ \cite{tag:27} indicating that the hexagons
are formed by the tilted-wave mechanism \cite{tag:28}, which is
the basic mechanism for resonator pattern formation \cite{tag:29}.
Dark-spot hexagons (FIG.~16~(a)) convert to bright-spot hexagons
(FIG.~16~(b)) when the driving intensity increases. This is in
qualitative agreement with numerical simulations (FIG.~10).

\begin{figure}[htbf]
\epsfxsize=70mm \centerline{\epsfbox{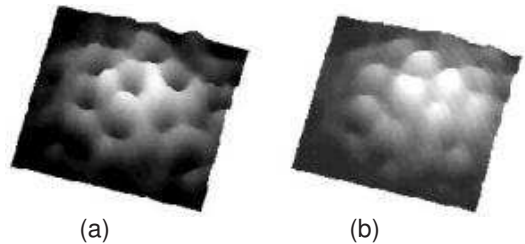}} \vspace{0.7cm}
\caption{(a) Bright (dark in reflection) and (b) dark (bright in
reflection) hexagonal patterns for the dispersive/defocusing case.
Driving light intensity increases from (a) to (b).}
\end{figure}

\begin{figure}[htbf]
\epsfxsize=80mm \centerline{\epsfbox{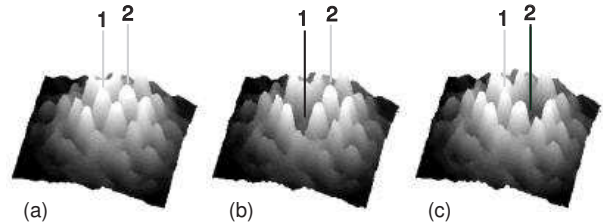}} \vspace{0.7cm}
\caption{Switching-off of individual spots of a hexagonal
structure with address pulses aimed at different bright spots
(marked 1 and 2) of the pattern.}
\end{figure}

\begin{figure}[htbf]
\epsfxsize=75mm \centerline{\epsfbox{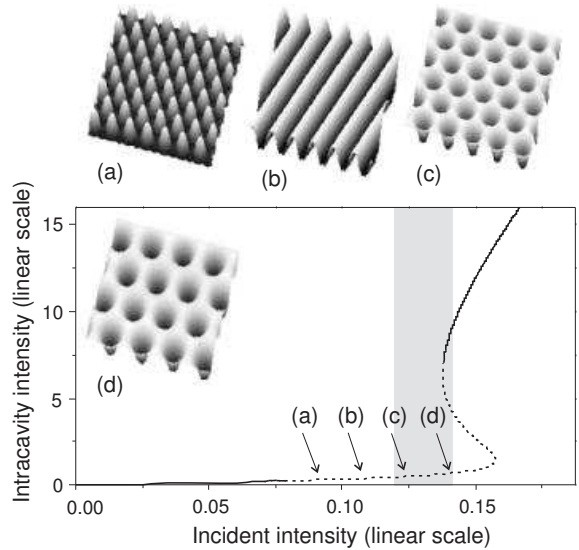}} \vspace{0.7cm}
\caption{Numerical solutions of Eq.(1) for intracavity light
intensity as function of incident intensity: homogeneous solution
(dashed line marks modulationally unstable part of the curve) and
patterns (a-d). Shaded area marks existence range for dark-spot
hexagons.}
\end{figure}

At high driving intensity we find that the bright spots in such
hexagonal patterns can be switched independently from one another
by focused optical (incoherent) pulses \cite{tag:27}. FIG.~17
shows the experimental results. FIG.~17~(a) shows the hexagonal
pattern formed. The focused light pulse can be aimed at
individual bright spots such as the ones marked "1" or "2".
FIG.~17~(b) shows that after the switching pulse was aimed at
"1", spot "1" was switched off. FIG.~17~(c) shows the same for
spot "2". We note that in these experiments we speak of true
logic switching: the spots remain switched off after the
switching pulse, (if the energy of the pulse is sufficient,
otherwise the bright spot reappears after the switching pulse).
These observations of local switching indicate that these
hexagonal patterns are not coherent patterns. The individual
spots are spatial solitons: they are independent, even at this
dense packing where the spot distance is about the spot size.

\begin{figure}[htbf]
\epsfxsize=80mm \centerline{\epsfbox{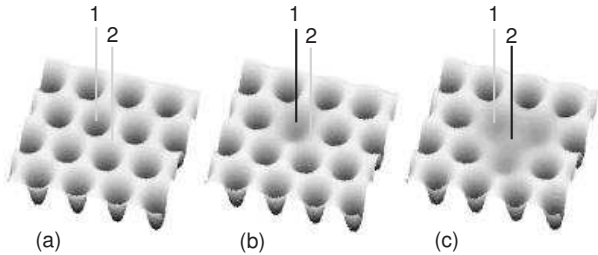}} \vspace{0.7cm}
\caption{Stable hexagonal arrangements of dark spatial solitons:
(a) without defects, (b) with single-soliton defect, (c) with
triple-soliton defect. Parameters as for FIG.~18 (d). Compare with
the experiment FIG.~17 and [27].}
\end{figure}

These experimental findings can be understood in the frame of the
model (1) \cite{tag:8}. FIG.~18 shows the bistable plane wave
characteristic of the semiconductor resonator for conditions
roughly corresponding to the experimental conditions. At the
intensities marked (a) to (d) patterned solutions exist.

The pattern period in FIG.~18~(a) scales again linearly with
$\theta^{-1/2}$ due to the tilted wave mechanism: When the driving
field is detuned, the resonance condition of the resonator cannot
be fulfilled by plane waves travelling exactly perpendicularly to
the mirror plane. However, the resonance condition can be
fulfilled if the wave plane is somewhat inclined with respect to
the mirror plane (the tilted wave mechanism \cite{tag:28}). The
system chooses therefore to support resonant, tilted waves.
FIG.~18~(a) is precisely the superposition of six tilted waves
that support each other by (nonlinear) 4-wave-mixing. The pattern
period corresponds to the resonator detuning as in the experiment
for structures FIG.~16~(a). In this pattern (FIG.~18~(a)) the
bright spots are not independent. Individual spots cannot be
switched as in the experiment FIG.~17.

On the high intensity pattern FIG.~18~(d) the pattern period is
remarkably different from FIG.~18~(a) even though the (external)
detuning is the same. This is indication that the internal
detuning is smaller and means that the resonator length is
nonlinearly changed by the intensity-dependent refractive index
(nonlinear resonance). From the ratio of the pattern periods of
FIG.~18~(a) and (d) one sees that the nonlinear change of
detuning is about half of the external detuning. That means the
nonlinear detuning is by no means a small effect. This in turn
indicates that by spatial variation of the resonator field
intensity the detuning can vary substantially in the resonator
cross section. In other words, the resonator has at the higher
intensity a rather wide freedom to (self-consistently) arrange
its field structure. One can expect that this would allow a large
number of possible stable patterns between which the system can
choose - or which are chosen by initial conditions.

FIG.~19 shows that at the high intensity corresponding to
FIG.~18~(d) the model (1) allows to reproduce the experimental
findings on switching individual bright spots. FIG.~19~(a) is the
regular hexagonal pattern, at high intensity. FIG.~19~(b) shows
the field with one bright spot switched off as a stable solution
and FIG.~19~(c) shows a triple of bright spots switched off as a
stable solution, just as observed in the experiments \cite{tag:8}.

Thus while FIG.~18~(a) is a completely coherent space filling
pattern, FIG.~18~(d) is really a cluster of (densest packed)
individual dark solitons. The increase of intensity from (a) to
(d) allows the transition from extended "coherent" patterns to
localized structures, by the increased nonlinearity, which gives
the system an additional internal degree of freedom. We note that
the transition from the coherent low intensity pattern to the
incoherent higher intensity structure proceeds through stripe
patterns as shown in FIG.~18~(b) \cite{tag:8}. For the intensity
of FIG.~18~(c) the individual spots are still not independent,
corresponding to the experiment FIG.~16~(b).

\subsubsection{Illumination near bandgap (absorptive/defocusing nonlinearity)}

Working at wavelengths close to the band edge we find bright and
dark solitons (FIG.~20), as well as collections of several spots
(FIG.~14~(b),(d),(e)) (all independent on the illumination beam
profile).

The nonlinearity of the MQW structure near the band edge is
predominantly absorptive. For comparing with calculations we can
therefore in the first approximation neglect the refractive part
of the complex nonlinearity in the model equations (1) and
describe the nonlinear medium as a saturable absorber. Numerical
simulations for this case (FIG.~21) confirm existence of both
bright and dark resonator solitons as they are observed in the
experiment (Figs~14,~20). We can contrast these purely dissipative
resonator spatial solitons with propagating spatial solitons (in a
bulk nonlinear material) \cite{tag:30}: the latter can not be
supported by saturable absorption.

\begin{figure}[htbf]
\epsfxsize=70mm \centerline{\epsfbox{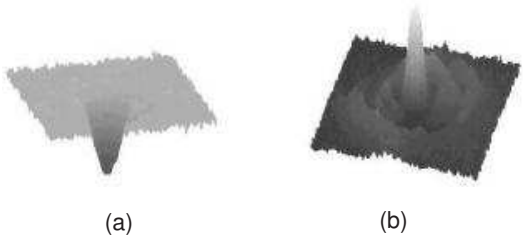}} \vspace{0.7cm}
\caption{Dark and bright solitons experimentally observed near
bandgap.}
\end{figure}

FIG.~22 shows details of the spontaneous formation of such bright
solitons as in FIG.~14~(c) and FIG.~20~(a). As discussed in
\cite{tag:24} material heating leads in this case to a slow
spontaneous formation of solitons, associated with the shift of
the semiconductor band edge by temperature \cite{tag:31}. In
FIG.~22 at ${t}$ $\approx$ 1.3 $\mu$s (arrow) the resonator
switches to high transmission (small reflection). The switched
area then contracts slowly to the stable structure FIG.~22~(b),
which is existing after ${t}$ $\approx$ 3.0 $\mu$s.

\begin{figure}[htbf]
\epsfxsize=70mm \centerline{\epsfbox{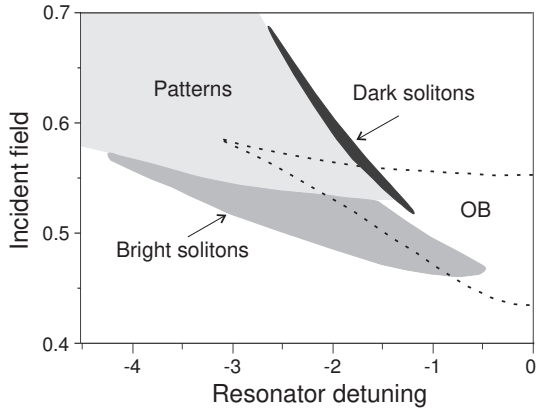}} \vspace{0.7cm}
\caption{Numerical solutions of Eq.(1) for unpumped ($P$=0 ),
absorptive (${\rm Re}(\alpha)=0$ ) case. Area limited by dashed
lines is optical bistability domain for plane waves. Shaded areas
are domains of existence of bright and dark solitons.  }
\end{figure}

After the resonator has switched to low reflection its internal
field and with it the dissipation is high. A rising temperature
$\Delta$$T$ decreases the band gap energy \cite{tag:31} ($E_{\rm
g}$ $\approx$ $E_{\rm go}$ - $\alpha$$\Delta$$T$) and therefore
shifts the bistable resonator characteristic towards higher
intensity. Thus the basin of attraction for solitons which is
located near the plane wave switch-off intensity (see locations of
the existence domains for the bright solitons and the plane wave
bistability in FIG.~21) is shifted to the incident intensity,
whereupon a soliton can form. Evidently for different parameters
the shift can be substantially larger or smaller than the width of
bistability loop, in which case no stable soliton can appear. In
consequence, we note that in absence of thermal effects (good
heat-sinking of sample) solitons would not appear spontaneously,
but would have to be switched on by local pulsed light injection.

\begin{figure}[htbf]
\epsfxsize=60mm \centerline{\epsfbox{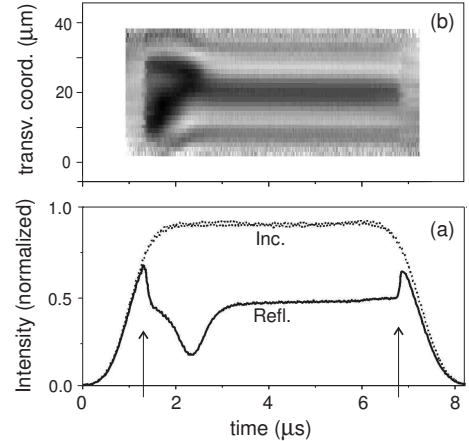}} \vspace{0.7cm}
\caption{Bright soliton formation below bandgap. Reflectivity on a
diameter of the illuminated area as a function of time (b).
Intensity of incident (dotted) and reflected (solid) light, at the
center of the soliton as a function of time (a). Arrows mark the
switch-on and -off. Details see text.}
\end{figure}

\begin{figure}[htbf]
\epsfxsize=60mm \centerline{\epsfbox{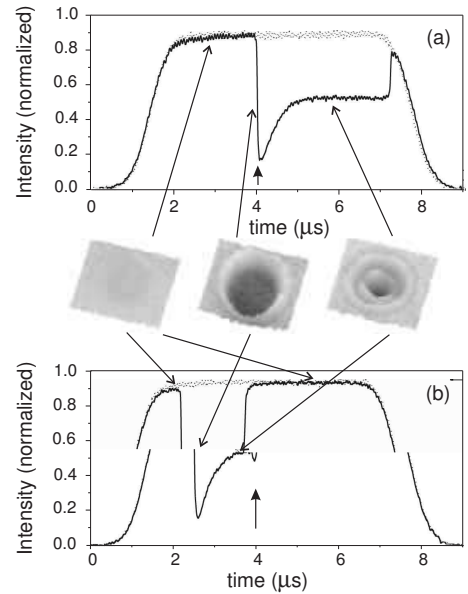}} \vspace{0.7cm}
\caption{Recording of incoherent switching-on (a) and
switching-off (b) of a soliton. Snapshot pictures show unswitched
state (left), circular switched domain (center) and a soliton
(right). Dotted trace: incident intensity, solid trace: reflected
intensity at center of soliton.}
\end{figure}

Besides the coherent switching (FIG.~15) bright solitons can also
be switched by light incoherent with the background field. Under
these conditions the switching occurs due to change of carrier
density alone. FIG.~23~(a) shows incoherent switch-on of a bright
soliton, where a perpendicularly polarized switching pulse of
$\approx$ 10 ns duration is applied at ${t}$ $\approx$ 4.0
$\mu$s. As apparent, a soliton forms after this incoherent light
pulse. The slow formation of the soliton is apparent in
FIG.~23~(a) (using roughly the time from ${t}$ $\approx$ 4.0
$\mu$s to ${t}$ $\approx$ 4.5 $\mu$s), indicating again the
influence of material heating.

It should be emphasized that this heating is not instrumental for
switching a soliton on. However, it allows switching a soliton
off incoherently \cite{tag:32}. This is shown in FIG.~23~(b)
where the driving light is initially raised to a level at which a
soliton forms spontaneously. The slow soliton formation due to the
heating is again apparent. The incoherent switching pulse is then
applied which leads to disappearance of the soliton.

\begin{figure}[htbf]
\epsfxsize=65mm \centerline{\epsfbox{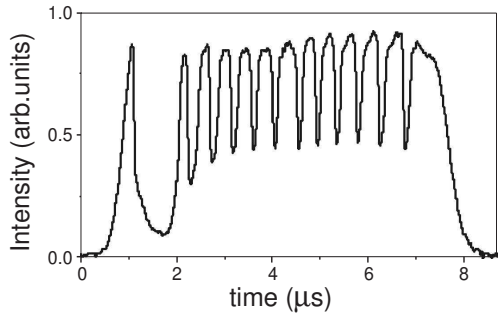}} \vspace{0.7cm}
\caption{Reflected light intensity measured at the center of a
soliton showing the regenerative pulsing (repeated switching on
and off of a soliton) resulting from combined thermal and
electronic effects.}
\end{figure}

\begin{figure}[htbf]
\epsfxsize=65mm \centerline{\epsfbox{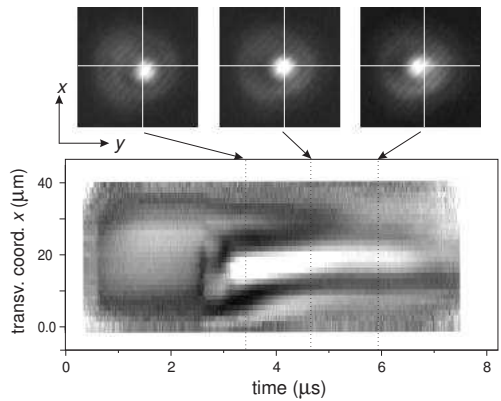}} \vspace{0.7cm}
\caption{Formation of a dark soliton (compare FIG.~22 and caption)
and motion. The motion visible in the streak picture (bottom) is
shown in 2D snapshots (top) for clarity. }
\end{figure}

The soliton can thus be switched on and also off by an incoherent
pulse. The reason for the latter is thermal effect: Initially the
material is "cold". A switching pulse leads then to the creation
of a soliton. Dissipation in the material at the location of the
soliton raises the temperature and the soliton is slowly formed.
At the raised temperature the band edge (and with it the
bistability characteristic and the existence range of solitons)
is shifted so that a new pulse brings the system out of the range
of existence of solitons. Consequently the soliton is switched
off.

Thus switching on a soliton is possible incoherently with the
"cold" material and switching off incoherently with the "heated"
material. When the driving intensity is chosen to be slightly
below the spontaneous switching threshold the nonlinear resonator
is cold. An incoherent pulse increases the carrier density
locally and can switch the soliton on, which causes local
heating. Another incoherent pulse aimed into the heated area can
then switch the soliton off and thereby return the resonator to
its initial temperature, so that the soliton could be switched
on/off again.

This thermal effect combined with electronic nonlinearity can
cause spatial and temporal instabilities. A bright soliton
switching on spontaneously in the cold material will then heat
locally and can thereby destroy the condition for its existence,
so that it switches off. After the material has cooled the
soliton switches on again etc. Regenerative pulsing of the
soliton results, an example of which is shown in the observation
FIG.~24.

As opposed to the bright soliton that heats the material locally
a dark soliton cools the material locally. This results in a
shift of the band edge and with it the switching characteristic
opposite to the bright soliton case. The consequence is that the
dark soliton moves laterally to places of uncooled material.
Here, however, it cools the material again so that a continuous
motion results, analogously to what we have described as the
"restless vortex" in \cite{tag:33}. Dark solitons in connection
with material heating effects tend therefore to be non
stationary. Such motion of a dark soliton was observed e.g. in
\cite{tag:24} and is shown in FIG.~25.

\subsubsection{Illumination above bandgap (absorptive/focusing case)}

At excitation above bandgap bright solitons form \cite{tag:34}
which have the same appearance as the bright solitons below/near
band gap (Figs~14~(c) and 20~(a)).

FIG.~26 shows the dynamics of the bright soliton formation for
excitation above bandgap. The difference between solitons
existing above and below bandgap can be understood from the
model. From (1) we obtain the reflected intensity as a function
of incident intensity for wavelengths above the bandgap(${\rm
Re}(\alpha)>0$), as well as below the bandgap (${\rm
Re}(\alpha)<0$), for plane waves (FIG.~27). One sees that the
bistability range is large below and small above the bandgap.
Solving (1) numerically, the typical bright soliton (top of
FIG.~27) is found coexisting with the homogeneous intensity
solutions in the shaded regions of FIG.~27~(a),~(b).

\begin{figure}[htbf]
\epsfxsize=60mm \centerline{\epsfbox{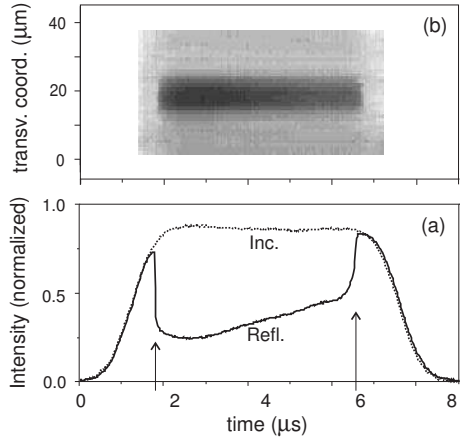}} \vspace{0.7cm}
\caption{Bright soliton formation above bandgap. Reflectivity on a
diameter of the illuminated area as a function of time (b).
Intensity of incident (dotted) and reflected (solid) light, at the
center of the soliton as a function of time (a). Arrows mark the
switch-on and -off. The soliton form fast (without mediation by a
thermal effect as in FIG.~22).}
\end{figure}

\begin{figure}[htbf]
\epsfxsize=85mm \centerline{\epsfbox{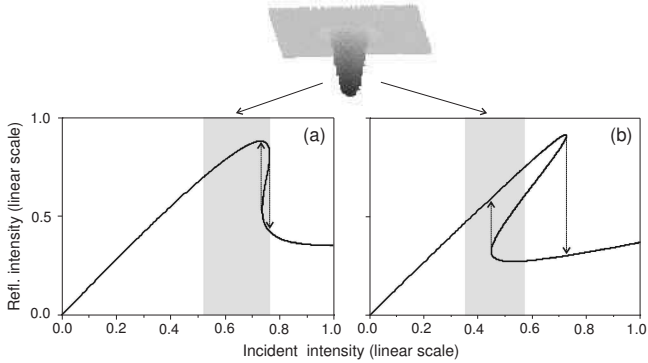}} \vspace{0.7cm}
\caption{Steady-state plane wave solution of Eq.(1) above bandgap
(a) and below bandgap (b). The soliton solution shown exists for
incident intensities corresponding to the shaded areas, in
coexistence with the homogeneous solution. For a temperature
increase the characteristics shift together with soliton existence
ranges to higher incident intensities. Reflected and incident
intensities normalized to the same value. Note the different
widths of the bistability ranges above and below bandgap.}
\end{figure}

Above the band gap FIG.~26 shows that the soliton is switched on
"immediately" without the slow thermal process described above.
FIG.~27~(a) shows why. The plane wave characteristic of the
resonator above band gap is either bistable in a very narrow
range, or even monostable (due to the contribution of the
self-focusing reactive nonlinearity \cite{tag:35}) but still with
bistability between the soliton state (not plane wave) and the
unswitched state. In this case the electronic switching leads
directly into the basis of attraction for solitons and the
switch-on of the soliton is purely electronic and fast. The widths
of the bistability characteristics observed experimentally
\cite{tag:34} correspond to the calculated ones FIG.~27.
Conversely the wide bistability range below bandgap requires the
(slow) thermal shift of the characteristic to reach the soliton
existence range.

Nonetheless, also above bandgap there is strong dissipation after
the switch-on. The associated temperature rise influences and can
even destabilize a soliton. The destabilization effect can be seen
in FIG.~26~(a). Over a time of a few µs after the soliton
switch-on the soliton weakens (reflectivity increases slowly)
presumably by the rise of temperature and the associated shift of
the band gap. At 6.5 $\mu$s the soliton switches off, although the
illumination has not yet dropped.

Thus, while the dissipation does not hinder the fast switch-on of
the soliton, it can finally destabilize the soliton. After the
soliton is switched off, the material cools and the band gap
shifts back so that the soliton can switch on again.

\subsubsection{Optical pumping}

The thermal effects discussed above result from the local heating
caused by the high intracavity intensity within the bright
soliton. They limit the switching speed of solitons and they will
also limit the speed at which solitons could be moved around,
thus limiting applications. The picture is that a soliton carries
with it a temperature profile, so that the temperature becomes a
dynamic and spatial variable influencing the soliton stability.

As opposed, a spatially uniform heating will not cause such
problems, as it shifts parameters but does not constitute a
variable in the system. The unwanted heating effects are directly
proportional to the light intensity sustaining a soliton. For
this reason and quite generally it is desirable to reduce the
light intensities required for sustaining solitons.

Conceptually this can be expected if part of the power sustaining
a resonator soliton could be provided incoherently with the
driving field, e.g. by means of optical or electrical pumping.
Pumping of the MQW structure generates carriers and allows
conversion from absorption to gain. If pumping is strong enough
the semiconductor microresonator can emit light as a laser 
\cite{tag:36}.

FIG.~28 shows the variation of the plane-wave bistability domain
with the pump as calculated from (1). The increase of the pump
intensity leads to a shrinking of the bistability domain for
plane waves (FIG.~28, pump intensities from $P_{\rm 1}$ to
$P_{\rm 3}$) and the resonator solitons' existence domain while
reducing the light intensity necessary to sustain the solitons.
This reduction of the sustaining light intensity was observed
experimentally (FIG.~29) \cite{tag:26}. When pumping below the
transparency point of the material and with the driving laser
wavelength near the semiconductor MQW structure band edge, bright
and dark solitons form similarly to the unpumped case (FIG.~20).
As the pump reduces the sustaining light intensity of the
solitons, the heating effects are weak and the solitons switch on
fast and unmediated by heating \cite{tag:26}.

\begin{figure}[htbf]
\epsfxsize=70mm \centerline{\epsfbox{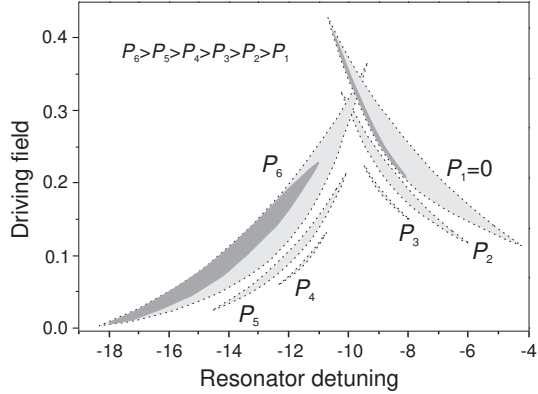}} \vspace{0.7cm}
\caption{Calculated plane-wave bistability domains as a function
of the pump intensity. Dark shaded areas show bright soliton
existence domains for two extreme cases: without pump ($P_{\rm
1}$=0) and with pump near lasing threshold ($P_{\rm 6}$).}
\end{figure}

\begin{figure}[htbf]
\epsfxsize=87mm \centerline{\epsfbox{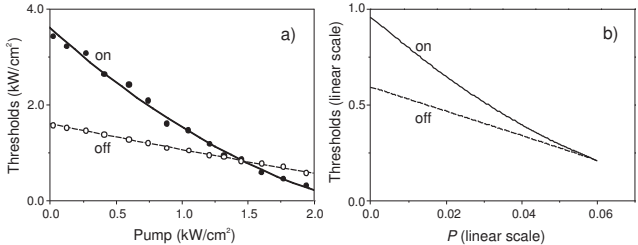}} \vspace{0.7cm}
\caption{Measured (a) and calculated (b) switch-on and switch-off
intensities for the driving light as a function of the pump
intensity. The unphysical crossing of the on and off curves is an
artifact from material heating.}
\end{figure}

When the pump intensity approaches the transparency point of the
semiconductor material, the resonator solitons' domain of
existence disappears (FIG.~28, pump intensities between $P_{\rm
3}$ and $P_{\rm 4}$). It reappears above the transparency point
(FIG.~28, pump intensities from $P_{\rm 4}$ to $P_{\rm 6}$) and
expands with the pump intensity (FIG.~30).

Switched structures observed below the lasing threshold are shown
in FIG.~31. FIG.~32 shows structures observed slightly above
lasing threshold when a driving field is used simultaneously with
the laser emission. The structures of FIG.~32 are reminiscent of
the solitons in electrically pumped resonators \cite{tag:37}. It
becomes clear that these structures (FIG.~32) must be
soliton-collections or patterns, and not (linear) mode patterns,
when looking at FIG.~32~(c). Here three bright spots are visibly
separated by darker lines. In a "mode-pattern" the phase change
from one bright spot to the next is $\pi$ (resulting in black
lines due to destructive interference separating the bright
spots). Therefore a "flower" mode pattern must have (and has)
always an even number of "petals", whereas here we observe an odd
number, inconsistent with phases in mode patterns. Thus we can
conclude that there is no phase change between the bright spots
and the latter are formed by self-localization, i.e. they are
solitons or patterns in a resonator above transparency of the
nonlinear material.

\begin{figure}[htbf]
\epsfxsize=70mm \centerline{\epsfbox{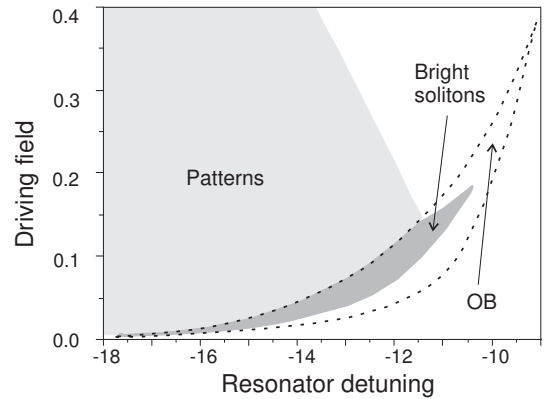}} \vspace{0.7cm}
\caption{Results of numerical simulations of below bandgap
solitons using the model (1) for a microresonator pumped close to
the lasing threshold. Shaded areas are domains of existence of
bright resonator solitons and patterns. Area limited by dashed
lines is optical bistability domain for plane waves. }
\end{figure}

We note that optically pumped resonators allow more homogeneous
pumping conditions than electrical pumping \cite{tag:38}. This
suggests that optical pumping lends itself more readily for
localization and motion control of solitons than electrical
pumping.

\begin{figure}[htbf]
\epsfxsize=80mm \centerline{\epsfbox{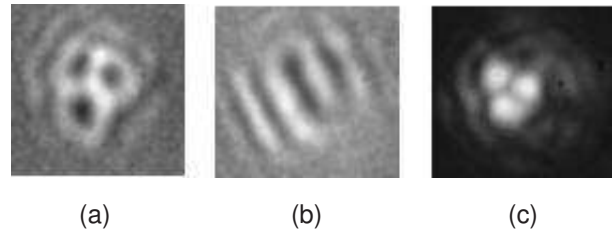}} \vspace{0.7cm}
\caption{Snapshots of typical optical structures at optical pump
intensities slightly below lasing threshold (driving light
intensity increases from (a) to (c)). With driving intensity dark
spot patterns change to bright spot patterns.}
\end{figure}

There is a difference between resonator solitons below bandgap in
pumped and unpumped material. The nonlinear resonance mechanism of
soliton formation \cite{tag:17} uses a defocusing nonlinearity
below transparency and a focusing nonlinearity above
transparency. Defocusing nonlinearity stabilizes dark solitons
and focusing nonlinearity stabilizes bright solitons. It follows
that dark solitons should prevail for unpumped material and
bright solitons for pumped material. FIG.~30 shows typical
examples of calculated resonator solitons for a pumped
semiconductor microresonator. Bright solitons have a large
existence range in the pumped case (FIG.~28, at $P_{\rm 6}$),
dark solitons exist, though with smaller range of stability, in
the unpumped case (FIG.~28, at $P_{\rm 1}$).

\begin{figure}[htbf]
\epsfxsize=80mm \centerline{\epsfbox{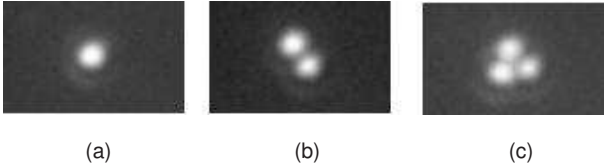}} \vspace{0.7cm}
\caption{Snapshots of typical optical structures at optical pump
intensities slightly above lasing threshold (driving light
intensity increases from (a) to (c)). Structures are always bright
spot patterns.}
\end{figure}

Thus pumped semiconductor resonators are well suited for
sustaining solitons below bandgap: (i) the background light
intensity necessary to sustain and switch resonator solitons is
substantially reduced by the pumping and therefore destabilizing
thermal effects are minimized, (ii) the nonlinear resonance
effect and the transverse nonlinear effect (self-focusing)
cooperate to stabilize bright solitons, therefore the domain of
existence of below bandgap (purely dispersive) bright solitons
can be quite large.

\section{Conclusion}

We have shown in these experiments that optical solitons
generally exist in nonlinear optical resonators. They can be
vortices, phase solitons, or bright and dark solitons. For
technical applications the experiments on semiconductor
microresonators have shown the existence of bright and dark
solitons. One can experimentally distinguish switched areas,
patterns and solitons from each other and from mode-fields (i.e.
fields whose structure is standing transverse waves resulting
from boundaries). Thermal effects can lead to spatial and
temporal instabilities. They can be controlled in various ways
e.g. by working above bandgap or by pumping. Pumping allows
furthermore to control the magnitude and the sign of the material
nonlinearities, and thus to maximize stability ranges for bright
(dark) solitons. Coherent periodic patterns can disintegrate with
increasing driving field into independent spatial solitons.

Optical resonator solitons show properties reminiscent of simple
biological structures and one may speculate that aggregations of
resonator solitons could allow information processing reminiscent
of brain functions.
\\

Acknowledgment\\ This work was supported by Deutsche
Forschungsgemeinschaft under grant We743/12-1.

\end{document}